\documentclass[doublecol]{epl2}

\usepackage{amsmath}

\title{A quasi-monomode guided atom-laser from an all-optical Bose-Einstein condensate}

\author{A. Couvert\inst{1} \and  M. Jeppesen\inst{1,2}, T. Kawalec\inst{1} \and  G. Reinaudi\inst{1} \and R. Mathevet\inst{3} \and  D. Gu\'ery-Odelin\inst{1,3}}
\shortauthor{A. Couvert \etal}

\institute{
\inst{1} Laboratoire Kastler Brossel, CNRS UMR 8852,
Ecole Normale
Sup\'erieure, 24 rue Lhomond, 75005 Paris, France\\
\inst{2} Australian Centre for Quantum Atom Optics, Physics
Department, The Australian National University, Canberra, 0200,
Australia\\ \inst{3} Laboratoire Collisions Agr\'egats
R\'eactivit\'e, CNRS UMR 5589, IRSAMC, Universit\'e Paul Sabatier,
118 Route de Narbonne, 31062 Toulouse CEDEX 4, France.}

\pacs{03.75.Pp}{Atom lasers} \pacs{03.75.Mn}{Multiple component
condensates; spinor condensate} \pacs{37.10.Gh}{Atom traps and
guides}

\abstract{We report the achievement of an optically guided and
quasi-monomode atom laser, in all spin projection states ($m_F =$
$-1$, $0$ and $+1$) of $F=1$ in rubidium 87. The atom laser source
is a Bose-Einstein condensate (BEC) in a crossed dipole trap,
purified to any one spin projection state by a spin-distillation
process applied during the evaporation to BEC. The atom laser is
outcoupled by an inhomogenous magnetic field, applied along the
waveguide axis. The mean excitation number in the transverse modes
is $\langle n \rangle = 0.65 \pm 0.05$ for $m_F = 0 $ and $
\langle n \rangle = 0.8 \pm 0.3$ for the low field seeker $m_F =
-1$. Using a simple thermodynamical model, we infer from our data
the population in each excited mode.}

\begin{document}

\maketitle

When atoms are coherently extracted from a Bose-Einstein
condensate (BEC) they form an atom laser, a coherent matter wave
in which many atoms occupy a single quantum mode. Atom lasers are
orders of magnitude brighter than
 thermal atom beams, and are first and second order coherent~\cite{BHE00,ORK05}. They are of
fundamental interest, for example, for studies of atom-light
entanglement, quantum correlations of massive
particles~\cite{HOH06} and quantum transport phenomena
\cite{LeP01,Pav02,CaL00,Car01,CEL02,PRS05,PLP05}. They are of
practical interest for matter-wave holography through the
engineering of their phase \cite{ODH00}, and for atom
interferometry because of their sensitivity to inertial
fields~\cite{Ber97}.

Many prospects for atom lasers depend upon a high degree of
control over the internal and external degrees of freedom and over
the flux. The control of the output flux in a pulsed or continuous
manner has been investigated using different outcoupling schemes:
short and intense radiofrequency pulses~\cite{MAK97}, gravity
induced tunneling~\cite{AnK98}, optical Raman pulses~\cite{HDK99},
long and weak radiofrequency fields~\cite{BHE99}, and by
decreasing the trap depth~\cite{CRG03}.

The control of their internal state is intimately related to the
outcoupling strategy. Atoms are either outcoupled in the
magnetically insensitive (to first order) Zeeman state $m_F = 0$
or another Zeeman state, each offering different advantages. Atom
lasers in $m_F = 0$ are ideal for precision measurement
\cite{Gil02} because of their low magnetic sensitivity. Atoms
in other Zeeman states, however, are ideal for measurements of
magnetic fields because of their high magnetic
sensitivity~\cite{VHL07}.

The control of the external degrees of freedom has been
investigated through the atom laser beam divergence while
propagating downwards due to gravity~\cite{RGL06,JDD08}. Inhomogeneous
magnetic field have been used to realize atom optical
elements~\cite{BKG01}. Recently, a guided and quasi-continuous
atom laser from a magnetically trapped BEC has been
reported~\cite{GRG06}.

In this Letter, we report on a new approach to generate guided
atom laser. This method can produce an atom laser in {\it any}
Zeeman state. In addition, our non-state changing outcoupling
scheme leads to an intrinsically good transverse mode-matching,
that enables the production of a {\it quasi-monomode} guided atom
laser. Therefore, we achieve simultaneously a high degree of
control of the internal and external degrees of freedom.

\begin{figure}
\onefigure[width=7.5cm]{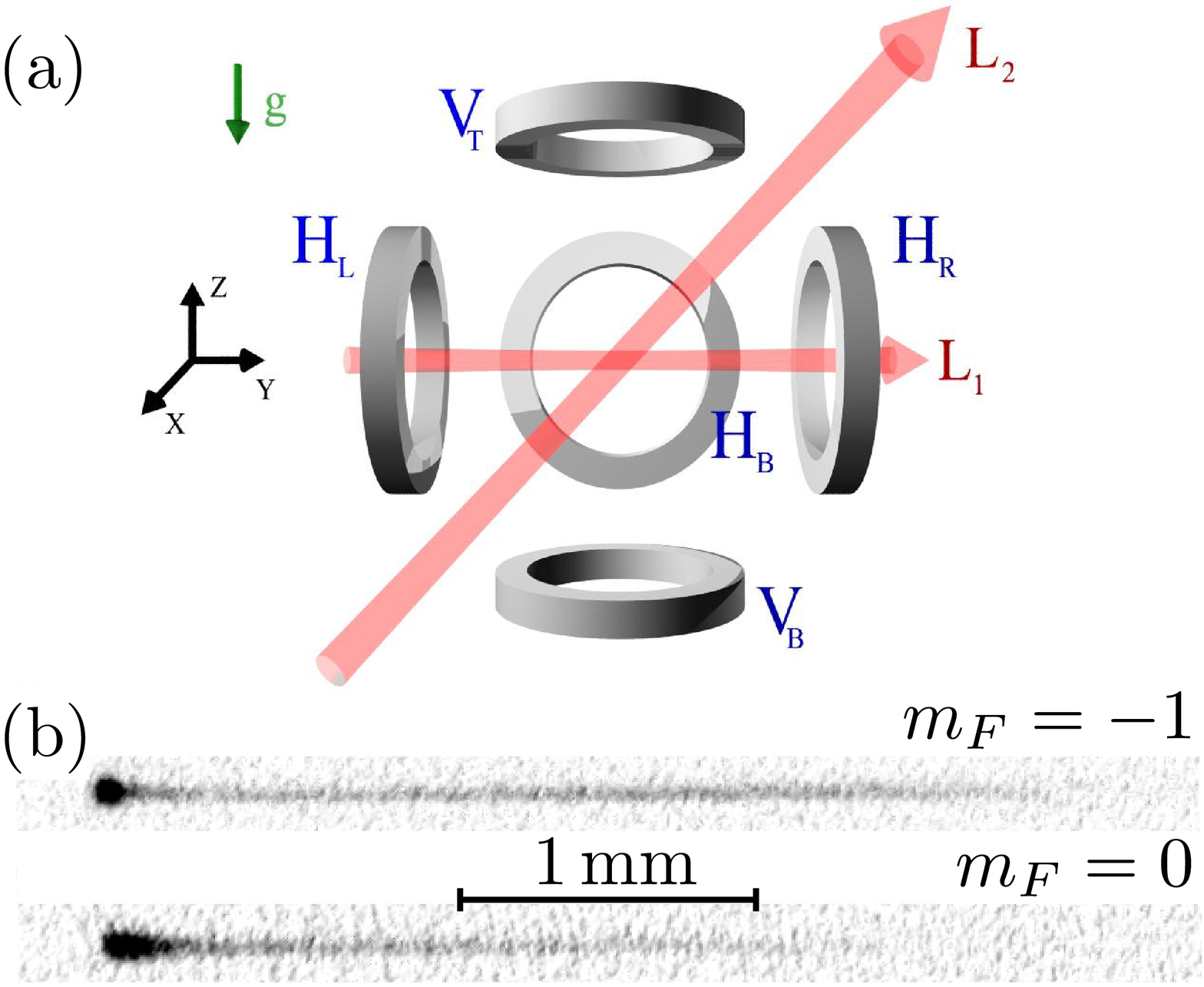} \caption{(Colour online). (a)
Schematic of experiment, showing trapping lasers and magnetic
coils.
 The cross dipole optical trap  is formed by two lasers of wavelength $\lambda = 1.07\, \mu$m,
  one horizontal ($L_1$) and one at $45^o$ ($L_2$). The coils are used individually to produce
  magnetic field gradients: during the evaporation ramp, the top coil $V_T$ is used for production
  of $m_F = +1$ condensates, the bottom coil $V_B$ for $m_F = -1$, and the off-axis horizontal coil $H_B$ for $m_F = 0$.
  Atom laser outcoupling is done with either on-axis coil, $H_L$ or $H_R$. (b) Absorption images for $m_F=-1$ and $m_F=0$
   atom lasers in waveguides, taken after a 15 ms expansion time.} \label{fig1}
\end{figure}

The atom laser is extracted from a Bose Einstein condensate
produced in a dipole trap. The trap is made from a Ytterbium fiber
laser (IPG LASER, model YLR-300-LP) with a central wavelength of
1072 nm and a FWHM linewidth of 4 nm. We have used an intersecting
beams configuration, formed by two focused linearly polarized
beams: beam ($L_1$) is horizontal and has a waist of $w_1\simeq
40\,\mu$m, and beam ($L_2$) which is in the $y-z$ plane at a 45
degree angle with respect to the horizontal beam, and a waist of
$w_2\simeq 150\,\mu$m (see fig.~\ref{fig1}(a)).

In order to control the laser power $P_i$ ($i=1,2$) of each beam,
we used the first-order diffracted beam from water-cooled
acousto-optic modulators, made in fused silica and designed for
high power lasers. The selected diffraction order ensures that
beam $(L_1)$ and $(L_2)$ have a frequency difference of 80 MHz. We
actively stabilized the pointing of the vertical beam, and
passively stabilized  the pointing of the horizontal beam to less
than 50 $\mu$rad.

 The experimental sequence begins by collecting around
$10^9$ atoms of $^{87}$Rb, in an elongated magneto-optical trap
(MOT), loaded from a Zeeman slower source in less than 2 seconds.
The elongated shape of the MOT results from the two-dimensional
magnetic field gradient configuration. The dipole trap is on
during the MOT loading, with powers $P_1=24$ W and $P_2=96$ W. To
maximize the loading of atoms into the dipole trap, the horizontal
beam $L_1$ is overlapped on the long axis of the MOT, and provides
a reservoir of cold atoms \cite{THK03}. In addition, we favor the
selection of atoms in the hyperfine level $5S_{1/2},F=1$ by
removing the repump light in the overlapping region, similar to
the dark MOT technique \cite{KDJ93}.

To evaporate, we switch off the MOT and we reduce the power in
each beam by typically two orders of magnitude, following the
procedures shown in Ref.~\cite{OGG01}, and we produce spinor
condensates of around $10^5$~atoms\footnote{For $m_F=0$
(respectively $m_F=-1$) spin state experiments, the horizontal
beam has a final power of 100 mW (respectively 85 mW) at the end
of the evaporation, and the vertical beam a final power of 9.77 W
(respectively 9.54 W).}. The entire experimental cycle is less
than six seconds.

To analyze the properties of the condensate, we use low-intensity
absorption imaging, with a variable time-of-flight (TOF), after
switching off the dipole beams. In addition, we can use the Stern
and Gerlach effect by applying a magnetic field gradient during
the expansion to spatially separate the spin components.

When evaporating with no magnetic field, we produce a condensate
with an approximately equal number of atoms in each $m_F$ spin
state. To produce the atom laser, we require a BEC of one pure
spin state. To do this, we use a single magnetic coil to produce a
gradient in the magnetic field amplitude $\nabla |\mathbf{B}|$,
and hence a force on
 the atoms due to the Zeeman effect. At the location of the atoms,
 this force is almost purely in one direction, along the axis of
 whichever coil is being used. We use such a force, perpendicular to the guide axis $(L_1)$, to produce a spin-polarized BEC of an arbitrary $m_F$
 state.

\begin{figure}
\onefigure[width=7.5cm]{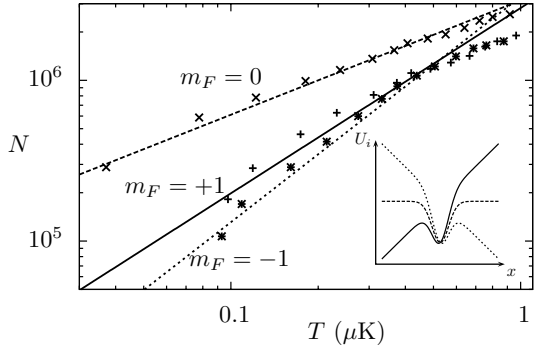} \caption{Spin distillation. Number
of atoms in each spin state versus temperature during the
evaporation
 process, showing the sympathetic cooling and purification of
 $m_F=0$ down to degeneracy. The purification is due to a horizontal magnetic field
 gradient, applied during the evaporation ramp. This field induces a force on $m_F=\pm
1$ that reduces the effective depth that they experience (see
inset, $U_i$ is the potential experienced by an atom in a $m_i$
Zeeman sublevel $(i=-1,0,+1)$).} \label{fig2}
\end{figure}

Spin distillation to one $m_F$ state occurs because, when the
force
 is applied during evaporation, the trap is less deep for the other
  $m_F$ states (see Inset of fig.~\ref{fig2}), and they are evaporated first~\cite{CRG03}.
   For example, to purify $m_F = 0 $, we use the horizontal coil $H_B$
   (see fig.~\ref{fig1}(a)), which forces the other $m_F$ states out of
   the trap, attracting the $m_F = +1$ and repelling the $m_F =-1$ \cite{LTA06}.
   To purify $m_F = -1$, we use the bottom coil $V_B$, which partially
   cancels the effect of gravity for $m_F = -1$, has little effect
    on $m_F =0$, and increases the effect of gravity for $m_F = +1$.
    Because atoms in the selected spin state are sympathetically cooled
     by the other atoms, the evaporation is more efficient, and we can produce
     condensates with a number of atoms in a given spin state approximatively three times greater than
     when evaporating with no field. We have confirmed that the three
  spin states are approximately at the same temperature during the entire evaporation.
In fig.~2, we show the evolution of each spin state population in
the course of evaporation for a spin distillation to the  $m_F =
0$ state. From the slope of the evaporation trajectories in the
number of atoms - temperature plot, we infer (using the
evaporation model of
  Ref.~\cite{OGG01}) that during the last second of evaporation, the ratio between trap depth
   and temperature $\eta$ is 5.1 for $m_F=0$ an 4.6 for $m_F= \pm 1$, which is further evidence for sympathetic cooling.

We demonstrate in the following that a BEC in a pure spin state
held in an optical trap can be coupled to the horizontal arm of
the trap in a very controlled manner, using magnetic forces along
the guide axis. The large Rayleigh length ($5$ mm for the
horizontal beam $L_1$) of the laser we use for the trap enables us
to guide the atom laser over several millimeters (see
fig.~\ref{fig1}(b)). The data reported on in this letter are for
the $m_F=0$ and the $m_F=-1$ spin states. Similar results have
been obtained with $m_F=+1$ state, using the same sequence as for
$m_F=-1$ but with the magnetic fields reversed.

After the evaporation is complete, we prepare to outcouple the
atom laser by linearly \emph{increasing} over 200 ms the power in
the horizontal beam to 200 mW for $m_F=\pm 1$ (resp. 400 mW for
$m_F=0$) and \emph{decreasing} the power in the vertical beam to
1~W for $m_F=\pm 1$ (resp. 800~mW for $m_F=0$).  This is done so
that the maximum available magnetic force will be sufficient to
outcouple. At the same time, we linearly increase a magnetic
gradient along the horizontal guide axis from 0 to 0.18~T/m, to
reach the threshold of outcoupling. Finally, to outcouple we hold
the power in each beam constant and increase the magnetic gradient
from 0.18~T/m to 0.22~T/m over a further 200 ms to generate the
beam. For atoms in $m_F = 0$ state, the force exerted by the
magnetic field is weaker than the one experienced by atoms in $m_F
= \pm 1$, since it relies on the second order Zeeman effect.
Nevertheless, we have demonstrated this magnetic outcoupling
method for all three spin states.

\begin{figure}
\onefigure[width=7.5cm]{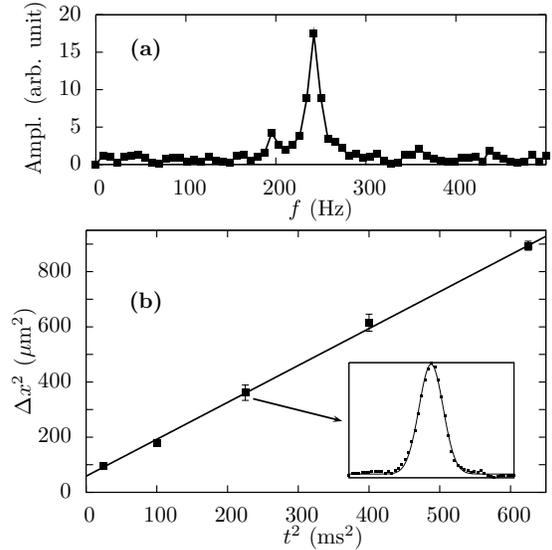} \caption{Measurements of the
trapping frequency and the velocity dispersion for rubidium atoms
in $|F=1,m_F=0\rangle$. (a) Fourier transform of atom cloud
oscillations in the waveguide. The peak has a frequency of 245 Hz
and a width of 10 Hz. (b) Plot of measurements of $(\Delta
x(t))^2$ against $t^2$ for the time of flight expansion of the
atom laser, and the straight line fit to the data, from which we
infer the velocity dispersion $\Delta v=1.2$ mm/s. Inset: Example
of Gaussian fit to integrated data, for a 15 ms TOF.} \label{fig3}
\end{figure}

The role played by the magnetic field is two fold: (i) it lowers
the trap depth along the optical guide axis and favors the
progressive spilling of atoms, and (ii) it accelerates the atoms
coupled into the guide, similar to gravity for standard non guided
atom lasers. Note that unlike gravity, we can control and even
turn off this inhomogeneous magnetic field at will.

The figure of merit for a guided atom beam is the number of
transverse modes that are populated. Experimentally, we cannot
measure directly those populations, but we have access to the mean
excitation number of the transverse modes $\langle n \rangle$.
Using the virial theorem for a non interacting beam\footnote{The
outcoupled beam is in the low density regime $\rho a \leq 0.1<1$,
where $a$ is the scattering length and $\rho$ the linear density
\cite{LeP01,Pav02}.} in a harmonic guide, one can show that the
transverse velocity dispersion $\Delta v$ obeys:
\begin{equation}
\frac{1}{2}m (\Delta v)^2 = \frac{\hbar \omega_z}{2} \left(
\langle n \rangle + \frac{1}{2} \right). \label{eq1}
\end{equation}
To infer $\langle n \rangle$, one needs to measure $\Delta v$ and
$\omega_z$. The trap frequency $\omega_z$ of the optical guide is
determined by observing the centre of mass oscillation of an atom
cloud prepared at this location. The transverse excitation is
produced by giving a momentum kick with a pulsed magnetic field.
fig.~\ref{fig3}(a) shows the Fourier transform of the oscillations
from which we infer a trap frequency of $\omega_z= 2\pi \times
(245 \pm 10)$ Hz. The same method gives a trap frequency of 170 Hz
for the experimental conditions used to investigate atom lasers in
the internal state $|F=1,m_{F} = 1\rangle$.

The velocity dispersion $\Delta v$ is inferred from a TOF
measurement. We integrate the absorption images along the atom
laser propagation direction over one millimeter \textit{i.e.} on a
distance over which the transverse frequency is constant to within
a few percent. The resulting one dimensional profiles are fitted
with a Gaussian function to find the width $\Delta x (t)$, $t$
being the TOF duration. We then find $ \Delta v$ through the
relation $ (\Delta x(t))^2 = (\Delta x_0)^2 + (\Delta v)^2 t^2$. A
typical
fit and TOF measurement can be seen in 
fig.~\ref{fig3}(b).

Using Eq.~\ref{eq1}, we find $\langle n \rangle = 0.8 \pm 0.3$ for
a guided atom laser of atoms in the $m_F=-1$ state, and $\langle n
\rangle = 0.65 \pm 0.05$ for the $m_F=0$ state\footnote{The larger
error bar for the $m_F=-1$ state might be attributed to a higher
sensitivity to technical noises due to the lower guide
frequencies.}. From our data, we conclude that this quality is
preserved as the beam propagates in the optical guide over several
millimeters. We have therefore produced a quasi-monomode guided
atom laser for any Zeeman state. Before analyzing in more detail
the transverse mode distribution,
 we discuss the issue of the flux of the laser beam.

\begin{figure}
\onefigure[width=8.5cm]{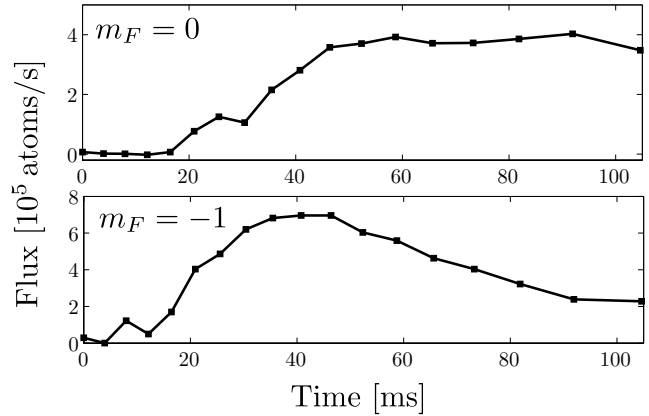} \caption{Flux of atom laser versus
time for the guided atom lasers depicted in fig.~\ref{fig1}(b), in
the $m_F=0$ and $m_F=-1$ spin states (a binning procedure has been
used to smooth out the noise due to the CCD camera).} \label{fig4}
\end{figure}

The control of the output flux is also a crucial issue for most
applications. Its variation as a function of time can be extracted
from the absorption images of fig.~\ref{fig1}(b). Those images are
taken after a 15 ms time of flight in absence of any magnetic
force. An atom that experiences the mean acceleration $\bar{a}$
($=\mu_Bb/2m$ for an $m_F=-1$ atom) where $b=20$ T/m during a time
$\tau$, is located after a time-of-flight of duration $t$ at
$y-y_0=\bar{a}\tau^2/2+\bar{a}\tau t$, where $y_0$ is the point of
outcoupling. From the absorption images we have a direct access to
the mean linear atomic density $\rho[y]$ along the propagation
axis. The flux is thus given by\footnote{The classical reasoning
performed here is justified since the length scales of interest
are much larger than the quantum length
$\ell=(\hbar^2/m^2\bar{a})^{1/3}<$ 1 $\mu$m on which the Airy-like
wave function oscillates.}: $\Phi (\tau)= \rho[y(\tau)]({\rm
d}y/{\rm d}\tau)$.
 We have shown in fig.~\ref{fig4} two examples of the
flux as a function of time, deduced from the measured density in
the absorption images of a guided atom laser in  $m_F=0$
 and in $m_F=-1$. For the $m_F=0$ state, the flux increases over the 30
first milliseconds to $4\times 10^5$ atoms/s, and then remains
constant over more than 70 ms. For the $m_F=-1$ state, a flux up
to $7\times 10^5$ atoms/s is reached in just 20 ms. The flux
decreases afterwards as the BEC get depleted.
 For our experimental sequence, the magnetic force is smaller for the
$m_F=0$ state compared to $m_F=-1$ state, and the smaller
outcoupling rate yields a nearly constant output flux for $m_F=0$.
During the outcoupling process, the flux is determined by the
chemical potential which is equal to the trap depth. Therefore, to
have the flux constant and stable over a long period of time would
require precise control of each beam's power and position, and of
the magnetic field gradient, all over the entire outcoupling.


The linear density $\rho$ extracted from our data ranges from
$5\times 10^7$ to $10^7$ atoms/m. We propose in the following a
simple model that gives the whole excitation spectrum for the
transverse degrees of freedom from the experimental values of the
linear density $\rho$ and the mean excitation number $\langle n
\rangle$. The beam is assumed to be a perfect Bose gas at
thermodynamical equilibrium made of $N$ atoms confined
longitudinally by a box of size $L=N/\rho$ with periodic boundary
conditions and transversally by a harmonic potential of angular
frequency $\omega$. The one-particle eigenstates of the system are
then labelled by three integers: the non-negative integers $n_X$
and $n_Z$ labelling the eigenstates of the harmonic oscillator
along the transverse $X$ and $Z$ axis, and the integer $\ell_Y$
labelling the momentum along $Y$. Replacing in the large $L$ limit
the sum over $\ell_Y$ by an integral, the normalization condition
reads:
\begin{equation}
 \rho\lambda = g_{1/2}(z) +
\sum_{p=1}^\infty \frac{z^p}{p^{1/2}} \left(
\frac{1}{(1-e^{-p\xi})^2}-1\right), \label{eq2}
\end{equation}
where we have introduced the de Broglie wavelength
$\lambda=\hbar/(2\pi m k_BT)^{1/2}$, the fugacity
$z=\exp(\beta\mu)$, the dimensionless parameter $\xi=\beta
\hbar\omega_z$ and the $p=1/2$ Bose function
$g_{1/2}(z)=\sum_{n=1}^\infty z^n/n^{1/2}$. Note that the function
$g_{1/2}(z)$ is not bounded when $z \rightarrow 1$, which means
physically that, in our trapping geometry, there is no Bose
Einstein condensation in the thermodynamical limit defined as $L,N
\rightarrow \infty$ with a fixed linear density $\rho=N/L$.
However, in such a combined box+harmonic confinement, Bose
Einstein condensation in the {\it transverse} ground state level
does occur at thermodynamical equilibrium \cite{MMD00}. The
expression for the critical temperature is :
\begin{equation}
T_c = \frac{\hbar \omega}{k_B} \left( \frac{\rho
\lambda}{\zeta(5/2)} \right)^{1/2}, \label{eq2bis}
\end{equation}
with $\zeta(5/2)\simeq 1.34$.

\begin{figure}
\onefigure[width=7.5cm]{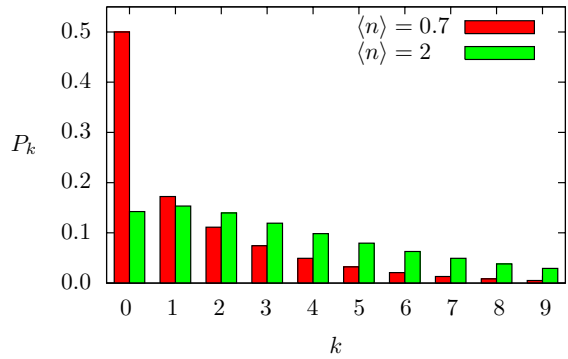} \caption{(Colour online). Transverse
excitation spectrum. Measuring the linear atomic density and the
mean excitation number $\langle n \rangle$, one can infer the
distribution of the transverse energy states
$\epsilon_k=k\hbar\omega$ using a thermodynamical equilibrium
approach where the atom laser beam is modelled by a Bose gas
confined by a box longitudinally and an harmonic potential
transversally. For a linear density equal to $5\times 10^7$
atoms/m, the occupation number of the ground state is on the order
of 50 \% for our data where we measure $\langle n \rangle\simeq
0.7$, and 14\% when $\langle n \rangle\simeq 2.0$ as reported in
Ref.~\cite{GRG06}.} \label{fig5}
\end{figure}

The mean excitation number $\langle n \rangle$ is obtained by
calculating the transverse energy $E_\perp=2N\hbar\omega\langle n
\rangle$. We find:
\begin{equation}
\langle n \rangle =  \frac{1}{\rho \lambda_0 \xi^{1/2}}
\sum_{p=1}^\infty \frac{z^p}{p^{1/2}}
\frac{e^{-p\xi}}{(1-e^{-p\xi})^3}, \label{eq3}
\end{equation}
with $\lambda_0=\hbar/(2\pi m \hbar\omega)^{1/2}$. From Eqs.
(\ref{eq2}) and (\ref{eq3}) and the experimental measurements of
$\langle n \rangle$ and $\rho$, we determine numerically the
dimensionless parameters $\xi$ and $z$ that allows one to infer
all the equilibrium properties of the Bose gas. As an example, the
occupation numbers $P_k$ of all transverse energy states
$\epsilon_k=k\hbar\omega$ are given by:
\begin{equation}
P_k = \frac{1}{\rho\lambda} (k+1)g_{1/2}\left(ze^{-k\xi} \right),
\label{eq4}
\end{equation}
where the $(k+1)$ factor accounts for the degeneracy of the state
$k$. For our data with a linear density equal to $\rho=5\times
10^7$ atoms/m and $\langle n \rangle\simeq 0.7$, we find 50 \% of
atoms in the transverse ground state and a temperature of 20 nK,
well below the critical temperature of 60 nK obtained for this
linear density according to Eq.~\ref{eq2bis}. The same calculation
yields 14\% of atoms in the transverse ground state for the
parameters
 and data reported in Ref.~\cite{GRG06} where $\langle n
\rangle\simeq 2.0$ (see Fig~\ref{fig5}). The thermodynamical
equilibrium assumption made in the previous reasoning is
approximately valid for the data presented in this letter. Indeed,
we have estimated that each atom undergoes a few collisions after
being outcoupled from the Bose Einstein condensate.

The atomic intensity fluctuations of the atom laser beam are
related to the dimensionless parameter
$\chi=\hbar^2\rho^3g/(mk_B^2T^2)$, where $g=g_{3D}/(2\pi a_0^2)$
with $g_{3D}=4\pi \hbar^2 a/m$, $a$ being the scattering length
and $a_0=(\hbar/m\omega_{z,x})^{1/2}$ \cite{Cas04}. For $\chi \gg
1$, small atomic intensity fluctuations are expected, and
conversely for $\chi \ll 1$. With our parameters, the decrease of
the atomic density as the atom laser propagates yields a decrease
of $\chi$ from $100$ to $1$ in $100$ ms. An interesting prospect,
with a better imaging system, deals with the study of atomic
intensity fluctuations and the investigation of the longitudinal
coherence with guided atom lasers produced in different
interacting regimes, including out-of-thermal equilibrium states.

Two effects are probably involved in the residual multimode
character of our atom laser. First, the residual thermal fraction
in equilibrium with condensate in the trap can populate the
excited transverse modes. Second, slight shaking of the position
of the relative position of beams $(L_1)$ and $(L_2)$ during the
outcoupling process can result in an increase value of the mean
excitation number. We therefore envision two main improvements for
future experiments: (i) a position locking scheme for the guide
with a bandwidth of a few kHz, and (ii) a more adiabatic
outcoupling by shaping the ramp of the magnetic field, which would
require a numerical optimization by solving the three-dimensional
Gross-Pitaevskii equation.

In conclusion, we have succeeded in smoothly coupling an optically
trapped BEC to a horizontal optical guide with low transverse
excitation. The near monomode nature of the atom laser will be
important in all applications which use phase engineering such as
matter-wave holography~\cite{FMS00}. A guided atom laser is an
ideal tool to investigate the transmission dynamics of coherent
matter waves through different structures. Such studies have their
counterpart in electronic transport phenomena, including the
generalization to cold atoms of Landauer's theory of
conductance~\cite{TWP99}, the atom-blockade
phenomenon~\cite{CaL00,Car01,CEL02}, non-linear resonant
transport~\cite{PRS05,PLP05}. Guided atom lasers in magnetically
sensitive states are ideal to combine with magnetic structures,
such as on atom-chips~\cite{Rei02}.

Finally, we emphasize that this work, combined with a continuous
replenishing of the optical dipole trap~\cite{CSL02}, can be
viewed as a promising strategy to generate a continuous guided
atom laser.

\acknowledgments

We thank C. Cohen-Tannoudji, J. Dalibard, Y. Castin and I.
Carusotto for useful comments and fruitful discussions.
 Support for this research came
from the D\'el\'egation G\'en\'erale pour l'Armement (DGA,
contract number 05-251487), the Institut Francilien de Recherche
sur les Atomes Froids (IFRAF) and the Plan-Pluri Formation (PPF)
devoted to the manipulation of cold atoms by powerful lasers.
G.~R. acknowledges support from the DGA, and M.~J. from IFRAF.

\end{document}